\documentclass[
    ,final            
  ]
  {aipproc}

\layoutstyle{6x9}
\usepackage{amsmath,amssymb}
\usepackage{bm}

\begin{document}

\title{Nucleon structure at large $x$: \\
	nuclear effects in deuterium}


\author{W. Melnitchouk}{
  address={Jefferson Lab, 12000 Jefferson Avenue,
		Newport News, Virginia 23606, USA}
}

\begin{abstract}
I review quark momentum distributions in the nucleon at large
momentum fractions $x$.
Particular attention is paid to the impact of nuclear effects in
deuterium on the $d/u$ quark distribution ratio as $x \to 1$.
A new global study of parton distributions, using less restrictive
kinematic cuts in $Q^2$ and $W^2$, finds strong suppression of the
$d$ quark distribution once nuclear corrections are accounted for.
\end{abstract}

\maketitle

\subsection{Introduction}

The momentum space distributions of quarks and gluons (partons) in
the nucleon provide fundamental characterizations of the nucleon's
bound state nature.
Considerable progress has been made in mapping out the parton
distribution functions (PDFs) of sea quarks and gluons in recent years
from deep inelastic scattering (DIS) and other high energy processes,
particularly at small values of the parton momentum fraction $x$.
In this region, however, the large fluctuation length of a virtual
photon into $q \bar q$ pairs means that it is not always clear
whether one is probing the structure of the target or the structure
of the probe itself.

At large values of $x$, where sea quarks and gluons play a negligible
role, the momentum distributions of valence quarks can be more
directly related to the nonperturbative structure of the nucleon.
The ratio of $d$ to $u$ quark distributions, for example, is very
sensitive to the mechanisms of spin-flavor symmetry breaking in the
nucleon \cite{NP}.
The large-$x$ region is also unique in allowing perturbative QCD
predictions to be realized for the $x$ dependence of PDFs in the
limit $x \to 1$ \cite{PQCD}.
Knowledge of PDFs at large $x$ is also important for searches of new
physics signals in collider experiments, where uncertainties in PDFs
at large $x$ and low $Q^2$ percolate through $Q^2$ evolution to affect
cross sections at smaller $x$ and larger $Q^2$ \cite{KUHLMANN},
as well as in neutrino oscillation experiments.

From high energy measurements involving proton targets one has
obtained a rather precise determination of the $u$ quark distribution,
which dominates the proton's valence structure due to its larger
charge weighting compared with the $d$.
Constraining the $d$ distribution, on the other hand, requires
in addition data on neutron structure functions.
However, because of the absence of free neutron targets, neutron
structure is usually extracted from a combination of deuteron and
proton data, which necessitates understanding of the nuclear
corrections in deuterium.
As a result, knowledge of PDFs at large $x$, and especially the $d$
quark distribution, has been severely limited beyond $x \sim 0.6$
\cite{CTEQ6X,ALEKHIN}.

In this talk, I will briefly review the status of nucleon structure
at large $x$, focusing in particular on nuclear effects in deuterium
and finite-$Q^2$ corrections, and present results from a new global
analysis \cite{CTEQ6X} which attempts to place stronger constraints
on large-$x$ PDFs.
My personal interest in large-$x$ physics began around 16 years ago
with a 1994 paper \cite{MSTOFF} with Tony Thomas and Andreas Schreiber
on DIS from off-shell nucleons.
It has been a pleasure to collaborate with Tony on this and many other
problems over the years.  I am also delighted to have Andreas, who has
since moved on to bigger and better things, present at this workshop.

\subsection{Nuclear effects in deuterium}

Because the deuteron is a very weakly bound nucleus, most analyses have
assumed that it can be treated as a sum of a free proton and neutron.
On the other hand, it has long been known from experiments on nuclei
that a nontrivial $x$ dependence exists for ratios of nuclear to
deuteron structure functions.
These effects include nuclear shadowing at small values of $x$
\cite{MT_SHAD}, anti-shadowing at intermediate $x$ values,
$x \sim 0.1$, a reduction in the structure function ratio below
unity for $0.3 \lesssim x \lesssim 0.7$, known as the European Muon
Collaboration (EMC) effect, and a rapid rise as $x \to 1$ due
to Fermi motion.

The conventional approach to describing nuclear structure functions
in the intermediate- and large-$x$ regions is the nuclear impulse
approximation, in which the virtual photon scatters incoherently
from the individual bound nucleons in the nucleus \cite{CONV}.
Furthermore, since quarks at large momentum fractions $x$ are most
likely to originate in nucleons carrying large momenta themselves,
the effects of relativity will be ever more important as $x \to 1$.
A relativistic description of the process therefore required the
development of a formalism for DIS from bound, off-shell nucleons,
which was pioneered in Ref.~\cite{MSTOFF}.
(Actually, the original motivation for that study was the quest
for a consistent description of pion cloud corrections to nucleon
PDFs, in particular the $\bar d/\bar u$ ratio, through the coupling
of the photon to an off-shell nucleon dressed by a pion \cite{MTV}.)

The off-shell DIS analysis \cite{MSTOFF} identified the conditions
under which usual convolution model \cite{CONV} of nuclear structure
functions holds, and found that in general these are not satisfied
within a relativistic framework.
In a follow-up study \cite{MST} (referred to as ``MST''), it was found
that one {\it can} however isolate a convolution component from the
total deuteron structure function, together with calculable off-shell
corrections.  The general expression for the deuteron $F_2$ structure
function can then be written as \cite{MST}
\begin{eqnarray}
F_2^d(x,Q^2)
&=& \sum_{N=p,n} \int dy\
    f_{N/d}(y,\gamma)\ F_2^N\left(\frac{x}{y},Q^2\right)\
 +\ \delta^{\rm (off)} F_2^d(x,Q^2)
\label{eq:F2d}
\end{eqnarray}
where $F_2^N$ is the nucleon structure function, and $f_{N/d}$ gives
the relativistic light-cone momentum distribution of nucleons in the
deuteron (also referred to as the nucleon ``smearing function'').
The scaling variable $y = (M_d/M) (p \cdot q/ p_d \cdot q)$ is
the deuteron's momentum fraction carried by the struck nucleon,
where $q$ is the virtual photon momentum, and $p (p_d)$ and
$M (M_d)$ are the nucleon (deuteron) four-momentum and mass.
In the Bjorken limit the distribution function $f_{N/d}$ is a
function of $y$ only and is limited to $y \leq M_d/M$.
At finite $Q^2$, however, it depends in addition on the ratio
$\gamma = |\bm{q}|/q_0 = \sqrt{1 + 4 x^2 M^2/Q^2}$ \cite{KP},
which can have significant consequences when fitting large-$x$
deuterium data \cite{KMK}.
Furthermore, at finite $Q^2$ the lower limit of the $y$ integration
is given by $y_{\rm min} = x (1 - 2M\varepsilon_d/Q^2)$, where
$\varepsilon_d$ is the deuteron binding energy, while the upper
limit is in principle unbounded \cite{CTEQX-2}.

The relativistic nucleon momentum distribution derived by MST \cite{MST}
(written here for simplicity in the $\gamma \to 1$ limit) is given by 
\begin{eqnarray}
f_{N/d}(y)
&=& { M_d \over 32 \pi^2 }\ y
    \int {dp^2 \over (M_d/E_p - 1)}\ \left| \Psi_d(\bm{p}) \right|^2\,
    \theta(p_0)\ ,
\label{eq:MST}
\end{eqnarray}
where $E_p = \sqrt{M^2 + \bm{p}^2}$ and $p_0 = M_d - E_p$
are the recoil and struck nucleon energies, respectively,
and $p^2 = p_0^2 - \bm{p}^2$ the struck nucleon's virtuality.
%
%
The deuteron wave function $\Psi_d(\bm{p})$ contains the usual
nonrelativistic $S$- and $D$-states, as well as the small $P$-state
contributions in relativistic treatments, and is normalized according
to
$\int d^3\bm{p} \left| \Psi_d(\bm{p}) \right|^2 / (2\pi)^3 = 1$.

Since the deuteron binding energy $\varepsilon_d = -2.2$~MeV is
$\approx 0.1\%$ of its mass and the typical nucleon momentum in the
deuteron is $|\bm{p}| \sim 130$~MeV, the average nucleon virtuality
$p^2$ will be $\sim 4\%$ smaller than the free nucleon mass.
For $x$ not too close to 1 one can therefore expanded the deuteron
scattering amplitude in powers of $\bm{p}/M$, using the so-called
weak binding approximation (WBA) \cite{KP,KMK,WBA}.
To order ${\cal O}(\bm{p}^2/M^2)$ one can then show explicitly that
the relativistic smearing function in Eq.~(\ref{eq:MST}) reduces to
the nonrelativistic WBA smearing function \cite{KP,KMK},
\begin{equation}
\hspace*{-0.3cm}
f_{N/d}(y)\
\overset{{\cal O}(\bm{p}^2/M^2)}{\longrightarrow}\
  \int {d^3p \over (2\pi)^3}\,
  \left( 1 + {p_z \over N} \right)\,
  \left| \Psi_D(\bm{p}) \right|
  \delta\left( y - 1 - {\varepsilon + p_z \over M} \right)\ \equiv\
f^{\rm WBA}_{N/d}(y)\, ,
\label{eq:WBA}   
\end{equation}
where $\varepsilon = M_d - M - E_p \approx \varepsilon_d - \bm{p}^2/2M$.
The resulting distribution function is sharply peaked around
$y \approx 1$, with the width determined by the amount of binding
(in the limit of zero binding it would be a $\delta$-function at $y=1$).
At finite $Q^2$ (or $\gamma$) the function becomes somewhat broader,
effectively giving rise to more smearing for larger $x$ or lower $Q^2$.

Finally, the convolution-breaking, off-shell correction
$\delta^{\rm (off)} F_2^d$ in Eq.~(\ref{eq:F2d}) receives
contributions from explicit $p^2$ dependence in the quark--nucleon
correlation functions, and from the relativistic $P$-state components
of the deuteron wave function.
This correction was estimated within a simple quark--spectator model
\cite{MST}, with the parameters fitted to proton and deuteron $F_2$
data, and leads to a reduction in $F_2^d$ of $\approx 1$ -- 2\% compared
to the on-shell approximation.

\begin{figure}
\includegraphics[height=.25\textheight]{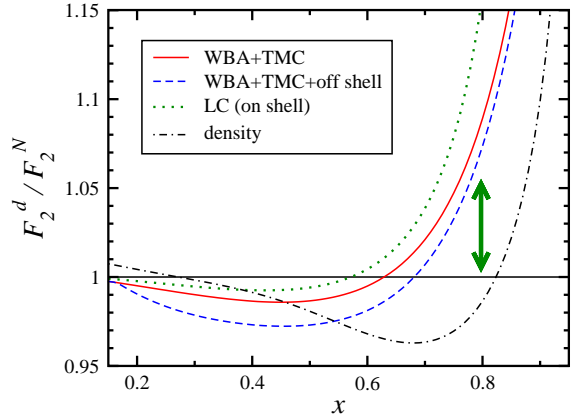}
\caption{Ratio of $F_2^d/F_2^N$ structure functions for the WBA
	smearing function with relativistic kinematics with
	(dashed) and without (solid) TMCs at $Q^2=5$~GeV$^2$.
	For comparison the ratio in the light-cone (dotted) and
	nuclear density extrapolation (dot-dashed) models are shown.}
\label{fig:RdN}
\end{figure}

The overall effect on the ratio $F_2^d/F_2^N$ is a $\sim 2$ -- 3\%
depletion relative to the free case at intermediate $x$ ($x \sim 0.5$),
with a steep rise at larger $x$ ($x \gtrsim 0.6$ -- 0.7) due to Fermi
motion, as illustrated in Fig.~\ref{fig:RdN} for $Q^2 = 5$~GeV$^2$.
Here the result for the WBA distribution (\ref{eq:WBA}), with
relativistic kinematics, is shown with and without the off-shell
correction from Ref.~\cite{MST}, and including finite-$Q^2$ target
mass corrections (TMCs) \cite{TMC}.
In both cases the EMC effect is larger than that obtained within a
light-cone approach \cite{FS81}, in which one assumes on-shell
kinematics and no binding.
The depletion at large $x$ is smaller, however, than that predicted
by the nuclear density extrapolation model \cite{FS88}, in which the
$F_2^d/F_2^N$ ratio is taken to scale with nuclear density.

Consequently, using binding/off-shell models one will extract a
{\em larger} neutron structure function from $F_2^d$ than with the
on-shell light-cone model \cite{NP}.
On the other hand, the extracted neutron will be {\em smaller} compared
to that obtained assuming the density model, or no nuclear effects at
all (see vertical arrows in Fig.~\ref{fig:RdN}).
Note that while a few global PDF analyses have attempted to incorporate
nuclear effects in the deuteron using smearing functions, most studies
simply neglect nuclear corrections altogether.
Although the extension of the density model to deuterium is problematic
\cite{COMMENT}, it is included here for reference since it is also used
sometimes to analyze deuterium data.

Finally, a word of caution against taking the ratios in
Fig.~\ref{fig:RdN} too literally.  From Eq.~(\ref{eq:F2d}) it is clear
that the deuteron structure function depends on both the details of the
nuclear physics embodied in $f_{N/d}$, and on the shape of the input
nucleon structure functions.
While the input proton structure function can be taken from experiment,
the neutron $F_2^n$ is unknown at large $x$, and generally a harder
$F_2^n$ input will lead to a larger EMC effect, pushing the rise of
$F_2^d/F_2^N$ above unity to larger $x$.
The practical solution is to perform an iteration procedure to
eliminate the dependence on the input $F_2^n$, or implement the
smearing directly in a global analysis, which is discussed next.

\subsection{New CTEQ6X distributions from large-$x$, low-$Q^2$ data}

Recently a global NLO analysis (referred to as ``CTEQ6X'') was performed
using an extended set of proton and deuteron data from DIS, from $pp$
and $pd$ Drell-Yan cross sections, $W^\pm$~asymmetry data, and jet cross
sections (see Ref.~\cite{CTEQ6X} for details).
The standard DIS cuts in previous global fits have excluded data with
$Q^2 < 4$~GeV$^2$ and $W^2 < 12.25$~GeV$^2$, effectively rendering
PDFs unconstrained above $x \approx 0.7$.
In the CTEQ6X fit the kinematical coverage was extended to larger $x$
by relaxing the $Q^2$ and $W^2$ cuts to $Q^2 > 1.69$~GeV$^2$ and
$W^2 > 3$~GeV$^2$, which approximately doubles the number of DIS
data points.

In any analysis of data extending into the low-$Q^2$ region, it is
imperative to account for kinematical target mass corrections
associated with finite values of $M^2/Q^2$ \cite{TMC}, as well as
dynamical $1/Q^2$-suppressed higher twist (HT) effects arising from
long distance multi-parton correlations.
For the CTEQ6X global analysis \cite{CTEQ6X} different prescriptions
for TMCs were considered, including the usual operator product expansion
approach, as well as a more recent formalism based on collinear
factorization.  For the HT correction a phenomenological parametrization
was applied, $F_2 = F_2^{\rm LT+TMC} (1 + C/Q^2)$, with the coefficient
$C$ determined empirically.
The fit was found to be stable with respect to the reduction of the
$Q^2$ and $W^2$ cuts, which is a rather nontrivial result given the
expanded kinematical coverage.
Remarkably, the leading twist PDFs turn out to be independent of
the TMC prescription adopted, {\em provided} the phenomenological
HT term is included.
This reveals an important interplay between the TMC and HT corrections,
which tend to compensate each other in the fitting procedure;
in contrast, without TMCs the HT alone cannot accommodate the full
$Q^2$ dependence of the data.

\begin{figure}[t]
\includegraphics[width=0.49\linewidth,bb=10 315 360 700,clip=true]{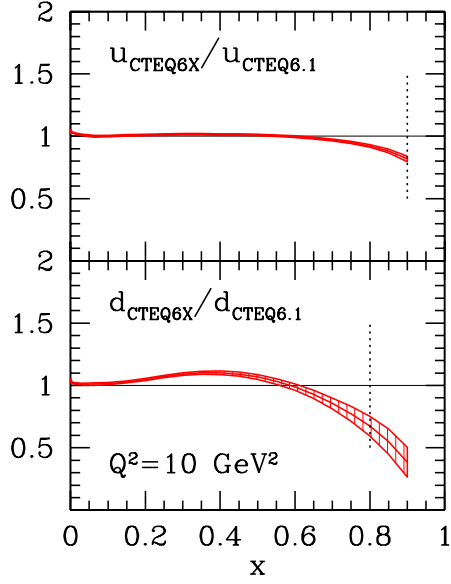}
\caption{CTEQ6X $u$ and $d$ distributions relative to the earlier
	CTEQ6.1 PDFs with no nuclear corrections \cite{CTEQ6X}.
	The vertical lines indicate the upper limits of validity
	of the fits.}
\label{fig:cteq6x}
\end{figure}

The inclusion of nuclear corrections in deuterium has profound effects
for the $d$ quark distribution.
Using the WBA finite-$Q^2$ smearing model, the $d$ distribution in the
CTEQ6X fit was found to be suppressed by up to 40\% for $x \approx 0.8$
relative to previous fits with no nuclear corrections, as
Fig.~\ref{fig:cteq6x} illustrates.
The $u$ distribution, which is strongly constrained by proton data,
is relatively unaffected by the nuclear corrections.
This trend is already clear from a comparison of the $F_2^d/F_2^N$ ratios
in Fig.~\ref{fig:RdN}, where the ratio in the nuclear smearing model is
$\gg 1$ at $x \approx 0.8$, so that the corresponding neutron $F_2^n$
structure function (and hence the $d$ distribution) will be smaller.
Not accounting for nuclear smearing in deuterium will therefore lead to
a significant overestimate of the $d$ distribution for $x \gtrsim 0.6$.
This will be the case for a wide range of nuclear smearing models, and
regardless of the details of the deuteron wave function.

The implication of a smaller $d/u$ ratio for nucleon structure is that
nonperturbative QCD physics, which generally predicts $d/u \to 0$ as
$x \to 1$ \cite{NP}, is still dominant at the currently accessible
$x$ and $Q^2$.
The behavior expected from perturbative QCD-inspired models, which
predict a finite $d/u$ ratio in the $x \to 1$ limit \cite{PQCD},
is not observed; whether this behavior will be revealed at even
larger $x$ remains to be seen.

\subsection{Outlook}

The fact that nuclear effects in deuterium play a vital role in
determining the structure of the neutron at large $x$ has been known
for some time.
As the focus of global PDF studies extends to larger values of $x$
and lower $Q^2$, with the availability of high-precision data from
Jefferson Lab and elsewhere, the need to incorporate deuterium
corrections is becoming paramount.
The CTEQ6X NLO fit has illustrated the significant impact of these
corrections on the $d$ quark distribution, which is found to be
suppressed by up to $\sim 40\%$ at the highest accessible value of $x$
($x \approx 0.8$) compared with earlier analyses with no nuclear effects.
Constraining the $d$ distribution at $x \gtrsim 0.8$ from inclusive
$F_2^d$ data will be challenging given the increasing uncertainty
in the nuclear corrections at larger nucleon momenta in the deuteron.

Further progress will be made with the help of several key
experiments planned at Jefferson Lab with 12~GeV.
This includes a novel idea of using the ratio of mirror symmetric
$^3$He and $^3$H nuclear structure functions, in which the nuclear
effects cancel to within $\sim 1\%$, to extract the $F_2^n/F_2^p$
ratio up to $x \approx 0.85$ \cite{A3}.
Another program already under way uses measurements of DIS on a
deuterium target with low-momentum spectator protons in the backward 
region to isolate an almost free neutron in the deuteron \cite{BONUS}.
Avoiding the use of nuclei altogether, yet another proposal utilizes
the weak interaction to probe the $d$ quark through parity-violating
electron DIS on a hydrogen target \cite{SOUDER,HOBBS}.
Here the asymmetry between left- and right-hand polarized electrons
selects the interference between $\gamma$ and $Z$-boson exchange,
which depends on the $d/u$ ratio weighted by electroweak charges,
and the expected 1\% asymmetry measurements would strongly constrain
$d/u$ up to $x \sim 0.8$ \cite{SOUDER}.

An exciting time lies ahead, with the expectation that the planned
program of measurements should finally close the book on one of the
longest-standing puzzles in the structure of the nucleon.


\begin{theacknowledgments}
I would like to thank my collaborators on the subject of nuclear effects
in the deuteron, including A.~W.~Schreiber, S.~Kulagin, A.~Accardi,
and most of all A.~W.~Thomas, who's been there from the beginning.
This work was supported by the DOE contract No. DE-AC05-06OR23177,
under which Jefferson Science Associates, LLC operates Jefferson Lab.
\end{theacknowledgments}

\end{document}